\documentclass[aps,amsmath,twocolumn,nofootinbib]{revtex4-2}
\usepackage{graphicx}
\usepackage{relsize}
\usepackage{enumitem}
\usepackage[table,usenames,dvipsnames]{xcolor}
\usepackage{ifthen}
\definecolor{linkcolor}{rgb}{0.6,0,0}
\definecolor{citecolor}{rgb}{0,0,0.75}
\definecolor{urlcolor}{rgb}{0.12,0.46,0.7}
\usepackage[breaklinks, colorlinks, urlcolor=urlcolor, linkcolor=linkcolor,citecolor=citecolor,pdfencoding=auto]{hyperref}
\hypersetup{linktocpage}
\setlist{nolistsep,leftmargin=*} %compress space for all lists

\begin{document}

%\voffset=0.75truein

\title{Deeper Learning in Astronomy}

\author{Douglas Scott} \email{docslugtoast@phas.ubc.ca}
\author{Ali Frolop} \email{afrolop@phas.ubc.ca}
\affiliation{Dept.\ of Physics \& Astronomy,
 University of British Columbia, Vancouver, Canada}

\date{1st April 2024}

\begin{abstract}
It is well known that the best way to understand astronomical data is
through machine learning, where a ``black box'' is set up, inside which
a kind of artificial intelligence learns how to interpret the features in
the data.  We suggest that perhaps there may be some merit to a new
approach in which humans are used instead of machines to understand the
data.  This may even apply to fields other than astronomy.
\end{abstract}

\maketitle

\section{Introduction}
One of us recently asked a computer what the best approach is for understanding
astronomical data.  The answer was, of course, ``machine learning''
\cite{machine}.  This comes as no surprise, since we are 
all well aware of the power and success of the traditional method of using
machines to comprehend data in astronomy, and indeed in other sciences.

There has been a long history of successful interpretation of data through
the use of machines and there are many good review papers on this
topic \cite{reviews}.
The general approach is to set up a kind of ``black box'', in which a
complex set of computations are carried out on existing data, in order to
extract the most salient features, interpret those features,
and perform analysis of new data -- in other words, to ``understand'' the
data.  The goal is to develop a machine that is
so complex and effective in its tasks that essentially an ``artificial
intelligence'' inside the machine is what is doing the understanding.

Machine learning is indisputably the best established means of solving problems
in astronomy, physics, and related fields.  Among its unquestionable
achievements, we now know that machine learning is the best way to search for
aliens \cite{aliens} and can be used to simulate the whole Universe
\cite{cosmology}.  In addition, artificial intelligence has been shown
to be smarter not only than a 5th grader, but also than physicists
such as Richard Feynman \cite{tegmark}.

Despite the obvious successes of this approach over a long period of time,
it may be worth exploring other ideas.
Indeed, the progress of science has shown that it is often useful to think of
alternative hypotheses, a technique sometimes called ``thinking outside
the box'' \cite{box}.  What we want to suggest here is a radical new scheme,
in which humans rather than machines are used to understand the data, something
we refer to as ``human learning'' or ``HL''.  While we appreciate that
this is never going to fully replace machine learning, we feel that there may
be situations where our proposal has some value, which we discuss in the rest
of this paper.

\section{Chatting}
For a particular context where HL might be useful, we have
found that people talking to each other can occasionally be helpful for
understanding difficult science issues.  We propose that general tools should
be developed to brainstorm topics and even to come up with whole sentences to
describe the ideas -- and we propose that these tools mostly use
{\it talking\/} to achieve
good results.  We think that such general tools should be called
``largely language methods'' or ``LLMs''.  One specific example
that we have been using ourselves is something we call
``Chat-over-Tea'' \cite{tea}.  The idea is that human beings gather to
drink hot beverages \cite{cookies,cakes} while having in-person discussions.
We have found that this is sometimes a fruitful way to tackle
complicated research questions \cite{allthetime}.

\section{Learning}
The use of graduate students to provide the labour for carrying out research
is well established.  Our HL idea can take this further, even to the extent
of encouraging meaningful interactions between graduate students and their
supervisors.  If we imagine a
university faculty member struggling to carry out novel research under the
weight of increasing bureaucracy, then it is natural that they become more and
more dependent on younger scientists to do their work for them \cite{latent}.
We like to
refer to each faculty member's group of students as the ``training set''.
When the students are new to the field, then the process is known as
``supervised learning''.  But once the students have gone through enough of
the HL process that they start to have their own ideas, then they reach the
level of ``unsupervised learning''.  It is definitely the case that use of
these ``graduate person units'' (or ``GPUs'') can improve the research
efficiency of senior scientists \cite{parallel}.
In fact there can sometimes be circumstances where the supervisor learns from
the junior researchers, a situation referred to as ``backpropagation''.

We suspect that somewhere in here there may be analogies
with how machine learning works, but we're not sure about this.

\section{Algorithms}
In more detail, there are many different approaches that one can take for
implementing HL.  Several specific examples suggest themselves.
Instead of using machine learning, there are some more esoteric methods we
would propose when trying to fit and interpret data, e.g.\ the use of
$\chi^2$ or ``maximum likelihood'' \cite{ML}; such methods might even be
useful when the data have Gaussian statistics.
If discussions among people get heated and groups break up into factions,
then we can have ``adversarial networks''; however, these are not as common as
they once were.  Where the discussions are less polarised, these group
interactions are called ``neutral networks'' or more specifically
``conversational neutral networks''.

There are many other techniques that might be worth implementing within HL.
In human discussions, splitting the interacting group into
``random threes'' could be useful.  When the session goes on for
too long, that's called the ``over-sitting problem''.
``Simplistic regression'' is a bit like striving for shallower learning.
Then there are TLAs \cite{TLAs}, such as SBI \cite{SBI}, SVM \cite{SVM},
KDE \cite{KDE}, SOM \cite{SOM}, etc.~\cite{further}.  These are all part of
the process of developing what we like to call ``non-artificial intelligence''.

\section{Big data}
Astronomy deals with space, space is big \cite{Douglas},
and hence astronomy involves
big amounts of data.
No one today can ignore the importance of Big Data.  We know this because
many tech experts have said so.  Pat Gelsinger (the C.E.O. of Intel)
said ``Data is the new science. Big Data holds the answers''.
Entrepreneur Douglas Merrill entreated that ``Big data isn't about
bits, it's about talent''.
And venture capitalist Chris Lynch ventured
``Big data is at the foundation of all of the megatrends that are happening
today, from social to mobile to the cloud to gaming.''
These experts may have a point, but they miss the most important issue,
which is that data are plural \cite{plural}.

\section{Conclusions}
Machine learning is clearly the ideal approach, to such an extent that it
has no disadvantages at all.  On the other hand, it seems only honest to
confess that there are some downsides inherent
in our proposed ``human-learning'' approach.
One is that it requires genuine effort on the part of at least some of the
humans involved.  Another criticism is that it may be hard to believe results
unless they come from a machine.  One more consequence is that the HL
approach may spoil claims of consciousness within artificial intelligence, by
revealing some of what is going on inside the
traditional machine-learning black box.  But despite these obvious drawbacks,
we feel that perhaps it might sometimes be a good thing for people to actually
understand what they're doing.

%\begin{acknowledgments}
\vspace{1.0cm}
\noindent{\bf Acknowledgements}:\ 
DS would like to thank many students and postdocs for making HL fun.
AF came out of retirement to write this paper \cite{us} and
will probably go straight back again; he would like to thank his old friend
Hermit Twain for suggestions.
No machines were harmed in the making of this paper.
%\end{acknowledgments}

%%%%%%%%%%%%%%%%%%%%%%%%%%%%%%%%%%%%%%%%%%%%%%%%%%%%%%%%%%%%%%%%%
%%%
%%%                     BIBLIOGRAPHY
%%%
%%%%%%%%%%%%%%%%%%%%%%%%%%%%%%%%%%%%%%%%%%%%%%%%%%%%%%%%%%%%%%%%%

\smallskip

%\newpage
%\vskip .75 in


\begin{thebibliography}{9}
\bibitem{machine}Actually, the computer wrote ``Certainly, here is a possible
answer to your question.  I'm a machine, so \dots''
\bibitem{allthetime}Which is a good thing, since one of us spends almost all
of their time in this pursuit.
\bibitem{latent}We might say that they drift into a ``latent space''.
\bibitem{ML}Also known as the ``ML'' method, to distinguish it from
machine learning.
\bibitem{plural}You might want to argue that while the word ``data'' is plural,
``big data'' should be treated as a singular noun phrase; however, there are
surely more ``data'' in ``big data'' than in ``data'' and hence it is
clearly plural.
\end{thebibliography}
\end{document}